%Paper: hep-ph/9310258
%From: UGOCCIONI@to.infn.it (Roberto Ugoccioni - tel. +39(11)670 7245)
%Date: Mon, 11 Oct 1993 9:40:41 +0100 (WET)

\magnification=1200
\tolerance=500
\hsize 6truein\hoffset 0.3175truecm
\vsize 8.5truein
%%%% local macros included here:
% innanzitutto si caricano i fonts nella dimensione 10punti
% (quelli in 12punti sono gia' caricati da plain.tex):
\font\eightrm=cmr10 at 10truept
\font\eighti=cmmi10 at 10truept
\font\eightsy=cmsy10 at 10truept
\font\eightbf=cmbx10 at 10truept
\font\eighttt=cmtt10 at 10truept
\font\eightit=cmti10 at 10truept
\font\eightsl=cmsl10 at 10truept
\font\sixrm=cmr7 at 7truept
\font\sixi=cmmi7 at 7truept
\font\sixsy=cmsy7 at 7truept
\font\sixbf=cmbx7 at 7truept
\font\fourrm=cmr5 at 5truept
\font\fouri=cmmi5 at 5truept
\font\foursy=cmsy5 at 5truept
\font\fourbf=cmbx5 at 5truept
% aggiustamento di un paio di parametri:
\skewchar\eighti='177   \skewchar\sixi='177   \skewchar\fouri='177
\skewchar\eightsy='60   \skewchar\sixsy='60   \skewchar\foursy='60
\hyphenchar\tentt=-1    \hyphenchar\eighttt=-1
% definiamo prima il passaggio al normale...
\def\tenpoint{\def\rm{\fam0\tenrm}%
  \textfont0=\tenrm \scriptfont0=\sevenrm \scriptscriptfont0=\fiverm
  \textfont1=\teni \scriptfont1=\seveni \scriptscriptfont1=\fivei
  \textfont2=\tensy \scriptfont2=\sevensy \scriptscriptfont2=\fivesy
  \textfont3=\tenex \scriptfont3=\tenex \scriptscriptfont3=\tenex
  \textfont\itfam=\tenit \def\it{\fam\itfam\tenit}%
  \textfont\slfam=\tensl \def\sl{\fam\slfam\tensl}%
  \textfont\ttfam=\tentt \def\tt{\fam\ttfam\tentt}%
  \textfont\bffam=\tenbf \scriptfont\bffam=\sevenbf
  \scriptscriptfont\bffam=\fivebf \def\bf{\fam\bffam\tenbf}%
%\tt \ttglue=.5em plus.25em minus.15em
\normalbaselineskip=12pt
\setbox\strutbox=\hbox{\vrule height8.5pt depth3.5pt width0pt}%
\normalbaselines\rm}
%
% ...poi definiamo il passaggio al piccolino...
\def\eightpoint{\def\rm{\fam0\eightrm}%
  \textfont0=\eightrm \scriptfont0=\sixrm \scriptscriptfont0=\fourrm
  \textfont1=\eighti \scriptfont1=\sixi \scriptscriptfont1=\fouri
  \textfont2=\eightsy \scriptfont2=\sixsy \scriptscriptfont2=\foursy
  \textfont3=\tenex \scriptfont3=\tenex \scriptscriptfont3=\tenex
  \textfont\itfam=\eightit \def\it{\fam\itfam\eightit}%
  \textfont\slfam=\eightsl \def\sl{\fam\slfam\eightsl}%
  \textfont\ttfam=\eighttt \def\tt{\fam\ttfam\eighttt}%
  \textfont\bffam=\eightbf \scriptfont\bffam=\sixbf
  \scriptscriptfont\bffam=\fourbf \def\bf{\fam\bffam\eightbf}%
%\tt \ttglue=.5em plus.25em minus.15em
\normalbaselineskip=10pt
\setbox\strutbox=\hbox{\vrule height7.5pt depth2.5pt width0pt}%
\normalbaselines\rm}
% ...e infine le parole magiche per cambiare taglia:
\def\tenmath{\tenpoint\fam-1 }
\def\eightmath{\eightpoint\fam-1 }

\def\wspnormbaseline{14truept}
\def\wsplittbaseline{12truept}
\baselineskip=\wspnormbaseline
\parindent 3truepc

\outer\def\beginsection#1\par{\medbreak\vskip 0.6truecm plus0.1cm minus0.1cm
      \message{#1}\leftline{\bf#1}\nobreak\vskip 0.4truecm plus0.1cm minus0.1cm
      \vskip-\parskip\nobreak\indent}
\outer\def\subsection#1\par{\medbreak\vskip 0.6truecm plus0.1cm minus0.1cm
      \message{#1}\leftline{\it#1}\nobreak\vskip 0.4truecm plus0.1cm minus0.1cm
      \vskip-\parskip\nobreak\indent}

\def\ref#1{$^{#1}$}             % use pseudo-references at first, then
                             % change to numbers.

\def\captionsize{\baselineskip \wsplittbaseline\eightmath\noindent}
\def\normalsize{\baselineskip \wspnormbaseline\tenmath}

\def\authorspace{\vskip 1.5truecm}
\def\abstractspace{\vskip 1.0truecm}
\def\abstract#1\par{\abstractspace
      \centerline{ABSTRACT}\vskip0.7truecm
      {\narrower\noindent #1\smallskip}}
\def\figure#1#2\par{\captionsize\noindent{\bf Figure #1} #2}
\def\table#1#2\par{\captionsize\noindent{\bf Table #1} #2}
\def\eqref#1{$#1$}
\def\figref#1{#1}
\def\tabref#1{#1}

\font\caps=cmcsc10              % small caps font (e.g. Jetset 6.3)
               % calligraphic P(Q_0,Q_1|Q)
          % integr.symbol with limits on top&bottom
\def\ee{$e^+e^-$}               % e+ e- (annihilations...)
             % GeV is roman inside mathmode

\def\nbar{\bar n}               % NB parameters: must be used in math mode
               % only, or it won't work!

  % <n>: requires math mode.
\def\jetset{{\caps Jetset}}

\def\luclus{{\caps Luclus}}

        %math mode only
\def\NF{{\cal N}_{\kern -1.9pt f}}     %math mode only
\def\NC{{\cal N}_{\kern -1.7pt c}}     %math mode only
       %math mode only

\def\roots#1{$\sqrt{s} = #1$ GeV}       % the well abused root s = xx GeV...
\def\pt{{p\kern -.2pt\lower 4pt\hbox{\fivei T}}}    %works well enough
                                                    %needs math mode
\def\ycut{y_{\rm cut}}          % math mode only
          % math mode only
\def\ptcut{\pt_{\rm cut}}
\def\pl{{p\kern -.2pt\lower 4pt\hbox{\fivei L}}}
          % requires math mode
\def\Dy{\Delta y}

%define less than or approx. / greater than or approx. - math mode
\def\nostrocostrutto#1\over#2{\mathrel{\mathop{\kern 0pt \rlap
  {\raise.2ex\hbox{$#1$}}}
  \lower.9ex\hbox{\kern-.190em $#2$}}}
   %less or around ...
   %greater or around...

\nopagenumbers
{
\baselineskip=12pt
\rightline{DFTT 41/93}             % number/year
\rightline{September 6, 1993}      % preprint date
}

\vskip 4truecm
\baselineskip 15pt
\centerline{\bf VOID ANALYSIS AND HIERARCHICAL}
\centerline{\bf STRUCTURE FOR SINGLE JETS}
\vskip 1truecm
\centerline{R. UGOCCIONI, A. GIOVANNINI and S. LUPIA}  % authors
\baselineskip 12pt
\vskip 0.4truecm
\centerline{\it Dipartimento di Fisica Teorica, Universit\`a di Torino and}
\centerline{\it INFN, Sezione di Torino, via Giuria 1, 10125 Torino, Italy}

{
\parindent 0cm
\footnote{}{\it Work supported in part by M.U.R.S.T.\
(Italy) under Grant 1992.}
}

\vfil
\midinsert
\baselineskip=12pt
\parindent 1.5truecm
\centerline{ABSTRACT}
{\narrower
\noindent
After a brief review of the theoretical basis of
void scaling function properties of hierarchical structure, we
analyze the phenomenological consequences at single jet level
in Monte Carlo \ee\ annihilation events. We find an interesting
alternative approach for characterizing quark and gluon jets.
\smallskip}
\endinsert

\vfil
\baselineskip 12pt
\centerline{Talk presented by R. Ugoccioni at the}
\centerline{Workshop ``Soft Physics and Fluctuations''}
\centerline{Cracow, Poland, 4--7 May 1993.}
\centerline{To be published in the Proceedings.}
\vfil
\eject

\pageno=1
\footline={\hfill\folio\hfill}
\def\D{\Delta}
\def\Dy{(\D y)}
\def\p0{P_0(\Delta y)}
\def\vi{{\cal{V}}(\Delta y)}
\def\v{{\cal V}}
\def\ljet{$l$-jet}
\def\hjet{$h$-jet}

\captionsize
\centerline{\bf VOID ANALYSIS AND HIERARCHICAL}
\centerline{\bf STRUCTURE FOR SINGLE JETS}
\authorspace
\centerline{R. UGOCCIONI*, A. GIOVANNINI and S. LUPIA}
\centerline{\it Dipartimento di Fisica Teorica, Universit\`a di Torino and}
\centerline{\it INFN, Sezione di Torino, via Giuria 1, 10125 Torino, Italy}
\abstract
After a brief review of the theoretical basis of
void scaling function properties of hierarchical structure, we
analyze the phenomenological consequences at single jet level
in Monte Carlo \ee\ annihilation events. We find an interesting
alternative approach for characterizing quark and gluon jets.

{\parindent0pt\footnote{}{\it Work supported in part by M.U.R.S.T.
(Italy) under Grant 1992}
\footnote{}{*\enspace Talk presented by R. Ugoccioni}}

\vskip 1cm
\null
\normalsize
%-------------------------------------------------------------------
\beginsection 1. Introduction

In a field like multiparticle dynamics, where the underlying theory
(QCD) is as yet too difficult to be useful for predictive calculations
at final particle level,
the study of experimental regularities and their interpretation is
of uttermost importance. As is widely known,
regularities which have been discovered at
complex levels include the NB regularity within final charged hadron
multiplicity distributions (MD's). More recently,
it was found experimentally \ref{1}
that separating the full sample of events
into subsamples with a fixed number of jets, NB regularity is satisfied
with better accuracy. Our studies of single jets\ref{2,3}
shows that the same regularity is reproduced at this very elementary
level.

In order to deepen our understanding of single jets MD's, we carried
out the analysis of the void structure in \ee\ annihilation
Monte Carlo events. The interesting result is that the differences
between quark and gluon jets seen in the clan analysis of MD's
\ref{2,3} appear relevant also in the `void' analysis,
which becomes also an alternative approach to characterizing
jets of different origin.

In Sec. 2 the present theoretical framework for the `void' function
is very briefly reviewed;
in Sec. 3 our results on Monte Carlo events are commented.

%-------------------------------------------------------------------
\beginsection 2. Hierarchical Correlation Functions and Void Scaling

The  void probability $P_0(\D y)$ is defined as
the probability of detecting no particles in the region of
phase space identified by the symbol $\D y$; here we make explicit reference
to a central rapidity interval, but it should be clear that any
other cut can be used: e.g., in Sec.~3 we will discuss the results
in rapidity and transverse momentum intervals; as another example,
in astrophysics $P_0$ is the probability
that a region of real space is empty of galaxies.\ref{4}

Knowing the void probability is formally equivalent to a knowledge
of the full MD, since it can easily be shown that
$$
P_n(\D y) = {(- \bar n)^n \over n!} {\partial^n \over \partial {\bar
n}^n} \p0.                                             \eqno(1)
$$
In practice however it is more useful to consider the
void function\ref{4}
$$\eqalign{
  \vi \equiv& - {1 \over \bar n(\D y)} \log \p0 \cr
  =& \sum\limits_{n=1}^{\infty} {\bigl( - \bar n(\D y) \bigr)^{n-1}
  \over n!} \kappa_n (\D y)\cr}                       \eqno(2)
$$
where $\kappa_n(\D y)$ is the reduced $n$-order cumulant in the
interval $\D y$ defined in the standard way as the $n$-fold integral of
the $n$-order reduced correlation function:
$$
\kappa_n(\D y) = \int_{\D y} dy_1\dots\int_{\D y} dy_n c_n(y_1\dots y_n).
                                                     \eqno(3)
$$
Notice that there is no average over different intervals of
the same size, contrary to what is sometimes done experimentally.
It will be seen in fact that the regularity is better satisfied
in central intervals, and degrades when moving away.

\topinsert
\captionsize
\parindent 0pt
{\bf Table 1.} The table shows the allowed shapes for a graphical
representation of correlation functions up to fourth order in
the Linked-Pair Ansatz (LPA), and in the Van Hove Ansatz (VHA).
\vskip 8.5truecm
\endinsert

Eq.~\eqref{3} is very helpful in connection with hierarchical
models: these assume that reduced correlation functions of any order can
be expressed as products of two-particle reduced correlation functions,
summed over all combinations of pairs,
$$
 c_n(y_1,\dots,y_n) =
  \sum\limits_{\alpha_t} A_{n,\alpha_t} \sum\limits_{\sigma}
  c_2(y_{i_1},y_{i_2}) \dots c_2(y_{i_{n-1}},y_{i_n})\;.
                                              \eqno(4)
$$
Here the two-particle reduced correlation functions $c_2(y_i,y_j)$,
linking particles of rapidity $y_i$ and $y_j$, are summed over all
non-symmetric relabelings $\sigma$ of the particles, and then summed over
all distinct topologies $\alpha_t$, with  weights $A_{n,\alpha_t}$
which depend only on the topology and not on energy or the phase space
interval considered. In Table~\tabref{1} the graphical representations
of two different hierarchical models are shown: edges correspond to
two-particle reduced correlation functions,
each line corresponds to a different topology,
and different graphs in the same line correspond to different
relabelings.
Two models are shown in Table~\tabref{1} which differ in the allowed
topologies:
the Linked Pair Ansatz\ref{5} (LPA) requires that a particle
appears at most twice in a given term of Eq.~\eqref{4}
(`snake' graphs in Table~\tabref{1}), which
then becomes simply a sum over permutations:
$$
c_n(y_1,\dots,y_n)\vert_{\rm LPA} =
A_n \sum\limits_{{\cal{P}}}
      c_2(y_{i_1},y_{i_2}) \dots c_2(y_{i_{n-1}},y_{i_n})
                                                    \eqno(5)
$$
On the other hand, the Van Hove Ansatz\ref{6} (VHA),
which was motivated by
the occurrence of NB regularity, allows in addition connections with three
particles (`star' diagrams in Table~\tabref{1});
in this case one can establish a recurrence relation
among correlation functions:
$$
c_n(y_1,\dots,y_n)\vert_{\rm VHA} =
  (n-1) \left\{  c_{n-1}(y_1,\dots,y_{n-1})c_2 (y_i,y_n)
   \right\}_{\rm symm.}
                                    \eqno(6)
$$
where the right-hand term is symmetrized over all particles.
Both these models however, coincide at the reduced cumulant level,
because integrating Eq.~\eqref{4} over a central rapidity interval
$\D y$ one finds the hierarchical structure for cumulants:
$$
\kappa_n(\D y) = A_n \kappa_2^{n-1}(\D y)         \eqno(7)
$$
It also worth noting that other, non-hierarchical models can
approximate relation \eqref{7}, thus making it impossible
to go backwards from the cumulant level to the correlation
functions level. One of these models, developed in the framework
of the $1/{\cal N}$ expansion, is examined in detail
in reference~7.

%-------------------------------------------------------------------
\beginsection 3. Monte Carlo Results

The considerations developed in the previous section will now be applied
to \ee\ annihilation events in different rapidity and
transverse momentum ($\pt$) intervals,
with focus on the analysis of single jet properties.

\ee\ annihilation events have been generated by using \jetset~7.2
(parton shower)\ref{8} with Lund string fragmentation as
hadronization prescription; the values of
the parameters of \jetset\ different from the default ones are listed in
reference 2. 2-, 3- and 4-jet events have been selected by
using the \luclus\ algorithm, and single jets have been identified
within each sample by the same algorithm;
for each sample 40000 events have been
considered at c.m. energies
\roots{91}, \roots{200}, \roots{500} and \roots{1000}.
The  analysis has been performed
in central rapidity intervals $|y| < \ycut$
with $\ycut$ from $0.25$ up to the kinematically available value and in
$\pt$ intervals $\pt < \ptcut$ with $\ptcut$ starting from 0.125~GeV/c.
Rapidity and $\pt$\ are defined with respect to the single jet axis.

\topinsert
\captionsize
\parindent 0pt
\vskip 9.7truecm
\vskip -0.3truecm
\settabs 2\columns\+\bf a&\bf b\cr
\vskip 0.3truecm
{\bf Figure 1.} Void function $\vi$ for single jet in \ee\
annihilation (JETSET 7.2) from {\bf a)} 2-jet events,
and {\bf b)} 3-jet events, as a
function of $\bar n \Dy \kappa_2 \Dy$  for different central rapidity
intervals $|y| \le \ycut$, with $\ycut$ = 0.25, 0.5, 1.0, 1.5, 2.0
at different c.m.\ energies.
The sample of single jets has been divided in 5 subsamples as indicated
in the figure. Vertical scale refers to the ``all $\pt$'' sample,
i.e., the subsample in which no cuts in $\pt$ have been applied.
Plots for subsequent subsamples are shifted down by 0.2.
Dashed curves show NB prediction.
\endinsert

\topinsert
\captionsize
\parindent 0pt
\vskip 9.7truecm
\vskip -0.3truecm
\settabs 2\columns\+\bf a&\bf b\cr
\vskip 0.3truecm
{\bf Figure 2.}
Same as in Fig.~1, but for {\bf a)} $h$-jets and {\bf b)} $l$-jets from
3-jet events. In the figure are indicated the jet energies
$E_J$ corresponding to the c.m.\ energies
\roots{91} (squares),  \roots{200} (circles),
\roots{500} (triangles) and  \roots{1000} (diamonds).

\endinsert

In Fig.~\figref{1a} we look at the sample of single jets obtained from
2-jet events: each event therefore contributes with 2 jets.
The figure shows the void scaling function $\v$ vs the product
$\nbar \kappa_2$.
Each point represents the value of the void function for particles
in a jet of given c.m.\ energy, given rapidity interval and given
$\pt$ interval.
For clarity, points referring to different $\pt$ intervals have been
plotted displaced from one another by a fixed amount. The dotted lines,
which represent NB behavior, become therefore superimposed when this shift is
removed.
Different center-of-mass energies are represented by
different symbols, and each
point refers to a different rapidity interval, up to $\ycut = 2.0$.
We notice firstly that there is a good scaling behavior, as
the dependence of the function $\v$
on energy, rapidity and transverse momentum is confined to its
dependence on the product $\nbar \kappa_2$. Secondly, we observe
there is good agreement with the dotted curve which represents NB behavior;
this was expected on the ground of our previous studies on MD's in
restricted domains of phase space\ref{2,3}.
As for large rapidity intervals, the observed violations  to the scaling,
which are not shown in this figure, are due to the lack of translational
invariance for the two-particle correlation function which invalidates
the hierarchical structure of cumulants.
Since 2-jet events are composed of a quark and an antiquark jet of the
same energy, one can conclude that quark jets fulfill NB regularity,
and $\v$-scaling behavior.

We now compare Fig.~\figref{1b}, where we show
the same quantities as in Fig.~\figref{1a}, but the single jets
here come from 3-jet events. Each event contributes to the sample with
a quark jet, an antiquark jet and a gluon jet, all of different energies.
In this figure, all contributions are superimposed. We see that
scaling behavior is less good than in the 2-jet sample (a fact
which also was expected from previous analysis), but still acceptable
for small $\pt$ intervals, and that
the data points are spread to larger values of $\nbar\kappa_2$ and
at smaller values of $\v$ than in the 2-jet sample.
Since 3-jet events contain a gluon jet, we can say that
void analysis seems sensitive
to the presence of a jet originated by a gluon. This is confirmed
by the analysis of 4-jet events~\ref{7} in which the spread
is even larger.

In order to specify better the role of the gluon jet in these events,
an energy scan of the 3-jet event sample was
performed~\ref{2,3}. From each event of the 3-jet sample the lowest (\ljet),
the
intermediate  and the highest (\hjet) energy jet have been
collected in three separated samples.
This separation is indeed suggested by perturbative QCD: according to
conventional wisdom, the behavior of \hjet\ samples is
expected to reflect that of quark (antiquark) jets and
the behavior of \ljet\ samples  that of gluon jets. The
jet energy dependence of the $h$- and $l$-jet samples has
been studied by further collecting
the jets in energy intervals 2 GeV wide.
The energy of each jet is the kinematically reconstructed energy.
For this program we analyzed 100000 3-jet events
at c.m. energies \roots{91}, \roots{200}, \roots{500} and \roots{1000}.

Figure \figref{2} shows the results of void analysis on the above sample.
In Fig.~\figref{2a} we see the \hjet\ sample, and Fig.~\figref{2b}
the \ljet\ sample.
The points are now labeled with jet-energy instead of c.m.\ energy, but
the same rapidity and $\pt$ intervals of the other figures have been used.
One finds a good scaling behavior and good agreement with expectation
of NB regularity at all energies, all $\ycut < 2.0$ and all $\ptcut$.
It should be noticed that these
two samples (\hjet\ and \ljet) are part of the samples that were shown
superimposed in Fig.~\figref{1b}. Isolating them and separating them in energy
has much improved their scaling behavior.
In Fig.~\figref{3} we compare the two samples, the \hjet\ and the \ljet\ at
the same jet-energy of 43-45 GeV (of course the samples come from events
of different c.m.\ energies).
The main feature here is the different spread of points, because
\hjet s appear to yield larger values of $\v$ than \ljet s.
One should remember that this implies that clans are larger in the case
of \ljet s.

In order to be assured that \hjet s are actually quark jets, one can
compare results from the 2-jet sample (where only quark jets appear)
and from the \hjet\ sample at the same jet-energy. It is apparent from
Fig.~\figref{4}
that the identification of \hjet s with quark jets can be made safely.
We have no pure gluon-jet sample with which to compare \ljet s.
However we noticed in Fig.~\figref{2b} that \ljet s also satisfy NB regularity,
and with different parameters from \hjet s. One expects that
if the \ljet\ sample were
a mixture of gluon and quark jets, i.e., a mixture of two NB like samples
with different parameters, NB regularity would not be
satisfied. Since on the contrary it is, we conclude that in the \ljet\ sample
the contamination of quark jets is negligible, and that we can continue
to treat \ljet s as gluon jets.

\topinsert
\captionsize
\parindent 0pt
\vskip 9.7truecm
\settabs 2\columns
\+\vtop{\hsize 2.8truein
{\bf Figure 3.}
Comparison of void function $\vi$ for $h$- and $l$-jets from 3-jet events as a
function of $\bar n \Dy \kappa_2 \Dy$  for the same rapidity
intervals as in Fig. 1 at
jet energy $E_J$ = 43-45 GeV ($h$-jets from c.m.\ energy
\roots{91}, $l$-jets from \roots{500}).
The sample of single jets has been divided in 5 subsamples as indicated
in the figure. Vertical scale refers to the ``all $\pt$'' sample. Plots for
subsequent subsamples are shifted down by 0.2.
Dashed curves show NB prediction.}
&%
\hbox to 0.2truein{\hfill}\vtop{\hsize 2.8truein {\bf Figure 4.}
Comparison of void function $\vi$ for single jets from
2-jet events and $h$-jets from 3-jet events as a
function of $\bar n \Dy \kappa_2 \Dy$  for the same rapidity
intervals as in Fig.~1
at jet energy $E_J$ = 43-45 GeV (from c.m.\ energy \roots{91}).
The sample of single jets has been divided in 5 subsamples as indicated
in the figure. Vertical scale refers to the ``all $\pt$'' sample.
Plots for subsequent subsamples are shifted down by 0.2.
Dashed curves show NB prediction.}\cr
\endinsert

We can therefore characterize \hjet s by a little amount of branching
which grows with energy (clans are small):
this situation is consistent with the
identification of the {\it \hjet\ as a quark (antiquark) jet within
which gluon bremsstrahlung is the dominant mechanism}.
For the \ljet\ sample branching plays a more relevant role than for the
\hjet\ one: data points spread along the NB curve showing more deviation from
Poissonian behavior. This result is consistent with the identification
of the {\it \ljet\ as a gluon jet within which gluon self-interaction is the
dominant mechanism}.

\pageinsert
\captionsize
\parindent 0pt
\vskip 8truecm
\vskip -0.3truecm
\settabs 2\columns\+\bf a&\bf b\cr
\vskip 0.3truecm
{\bf Figure 5.}
{\bf a)} Single particle inclusive rapidity distributions $dn/dy$
vs. $y$ for $h$-jets from 3-jet events at $E_J$ = 43-45 GeV from \roots{91}
(solid line) and $E_J$ = 248-250 GeV from \roots{500} (dashed line);
rapidity is defined in the ancestor frame. Plots are normalized to
average multiplicity $\bar n = 8.8 $ at c.m.\ energy \roots{91}
and $\bar n = 14.5$  at \roots{500}.
{\bf b)} Same as in (a) but for $l$-jets from 3-jet
events at $E_J$ = 8-10 GeV from c.m.\ energy \roots{91} (solid line)
and $E_J$ = 43-45 GeV from  \roots{500} (dashed line).
Plots are normalized to
average multiplicity $\bar n = 7.7$ at \roots{91} and $\bar n = 11.8$
at \roots{500}.

\vskip 7truecm plus 2cm
\vskip -0.3truecm
\settabs 2\columns\+\bf \hphantom{XXXXXX}a&\bf \hphantom{XXXXXX}b\cr
\vskip 0.3truecm
{\bf Figure 6.}
{\bf a)}\ Single particle normalized inclusive rapidity and
$\pt$\
distributions  $1 / \bar n ( {d^2n/dy d\pt})$ vs. $(y,\pt)$
for $h$-jets from 3-jet events at jet energy $E_J$ = 43-45
GeV from c.m.\ energy \roots{91};
rapidity is defined in the c.m.\ frame.
{\bf b)}\ Same as in (a) but for $l$-jets from 3-jet events
at jet energy $E_J$ = 43-45 GeV from \roots{500}.

\endinsert

Further support to this result comes from the study of single particle
inclusive rapidity and transverse momentum distributions separately
for \hjet s and for \ljet s; for this program 50000 events have been
generated. The inclusive rapidity distributions $dn/dy$ are shown
vs rapidity $y$ in Fig.~\figref{5} at different jet energies: rapidity
is with respect to the ancestor of each jet, event by event; distributions
are normalized to the average multiplicity. The distribution for \ljet s
(Fig.~\figref{5b}) is peaked at $y=0$, while the distribution for
\hjet s (Fig.~\figref{5a}) is more flat.
Both distributions grow slowly with jet energy.
The inclusive $(y,\pt)$ distributions are shown in Fig.s~\figref{6a}
(\hjet s) and \figref{6b} (\ljet s): here the distributions
can be compared directly, because both have
the same jet energy ($E_J$ = 43-45 GeV),  both are normalized to 1
and rapidity is taken with respect to the c.m.\ frame of the annihilation.
The \ljet\ is seen to be
less extended in rapidity but more spread out in $\pt$, as generally
expected for a gluon jet with respect to a quark jet.

\topinsert
\captionsize
\parindent 0pt
\vskip 9truecm
{\bf Figure 7.}
Average distance in rapidity between particles for $h$- and
$l$-jets from 3-jet events as a function of jet energy; dotted less
dense area shows $h$-jets, dotted more dense area shows $l$-jets.

\endinsert

In conclusion, all results on single particle inclusive distributions
support the idea that particles inside a gluon jet, where branching
plays a relevant r\^ole, show more correlation than in a quark jet, where
gluon-bremsstrahlung emission is dominant.
This picture is strengthened finally by Fig.~\figref{7}, where the average
distance in rapidity between particles is plotted against jet energy,
separately for \hjet s, which lie all within the less dense area,
and for \ljet s, which lie all within the more dense area. \ljet s
show more concentration in rapidity (and therefore more correlation)
than \hjet s, as particles inside a gluon jet are closer together
than particles inside a quark jet, and in agreement with the fact
that the aggregation parameter (NB parameter $k^{-1}$)\ref{9}
is larger for \ljet s than for \hjet s.

\vfill

%-------------------------------------------------------------------
\beginsection 4. Conclusions

The analysis of the structure of voids has been presented as a
tool to explore the hierarchical structure of correlations in
multiparticle dynamics.
Its application to samples of single jets obtained from \ee\
annihilation Monte Carlo events has shown that
quark and gluon jets can be characterized by
means of the respective void properties.
It should be noticed that the results obtained are consistent with
the presence of NB regularity {\it at single jet level},
and suggest a hierarchical
structure for correlations within single jets.

%-------------------------------------------------------------------
\beginsection 5. Acknowledgments

One of us (R.U.) would like to thank prof.\ K.~Fia{\l}kowski and
prof.\ A.~Bia{\l}as, and the Local Organizing Committee,
for the fruitful and stimulating atmosphere
that was created at this meeting.

%-------------------------------------------------------------------
\def\name{\item}
\beginsection 6. References

\vskip-\parskip

\name{1.}
P.\ Abreu et al., DELPHI Collaboration,
{\it Z.\ Phys.\ }{\bf C56} (1992) 63

\name{2.}
F. Bianchi, A. Giovannini, S. Lupia and R. Ugoccioni,
{\it Z. Phys.\ }{\bf C58} (1993) 71

\name{3.}
F. Bianchi, A. Giovannini, S. Lupia and R. Ugoccioni,
``Single Jet Multiplicity Distributions in $e^+e^-$ Annihilation at
High Energies'', DFTT 49/92, to be published in the {\it Proceedings of the
XXII International Symposium on Multiparticle Dynamics},
Santiago de Compostela, Spain, 13--17 July 1992

\name{4.}
J.N. Fry, {\it Ap.\ J.\ }{\bf 306} (1986) 358

\name{5.}
P.\ Carruthers and I.\ Sarcevic,
{\it Phys.\ Rev.\ Lett.\ }{\bf 63} (1989) 1562

\name{6.}
L. Van Hove, {\it Phys.\ Lett.\ }{\bf B242} (1990) 485

\name{7.}
S. Lupia, A. Giovannini and R. Ugoccioni,
{\it Z. Phys.\ }{\bf C59} (1993) 427

\name{8.}
M.\ Bengstsson and T.\ Sj\"ostrand, {\it Nucl.\ Phys.\ }{\bf B289} (1987) 810;
\hfill\break
T.\ Sj\"ostrand and M.\ Bengstsson, {\it Computer
Physics Commun.\ }{\bf 43} (1987) 367

\name{9.}
A. Giovannini and L. Van Hove, {\it Z.\ Phys.\ }{\bf C30} (1986) 391

\bye